\documentclass{PoS}

\usepackage[numbers]{natbib}
\usepackage{url}

\title{Parsec scale polarization properties of the TeV blazar Markarian 421}

\ShortTitle{Polarization analysis of Mrk~421 at 15 and 24 and 43 GHz during 2011}

       \author{\speaker{R.~Lico}$^{,1,2}$, M.~Giroletti$^{1}$, M.~Orienti$^{1,2}$, J.~L.~G\'omez$^{3}$,  C.~Casadio$^{3}$, F.~D'Ammando$^{1,2}$, M.~G.~Blasi$^{1,4}$, W.~Cotton$^{5}$, P.~G.~Edwards$^{6}$,  L.~Fuhrmann$^{7}$, S.~Jorstad$^{8,9}$, M.~Kino$^{10}$, Y.~Y.~Kovalev$^{11,7}$,  T.~P.~Krichbaum$^{7}$,  A.~P.~Marscher$^{8}$, D.~Paneque$^{12}$, B.~G.~Piner$^{13}$, K.~V.~Sokolovsky$^{11,14}$. \\
              
$^{1}$INAF Istituto di Radioastronomia, Bologna, Italy. \\
$^{2}$Dipartimento di Fisica e Astronomia, Universit\`a di Bologna, Italy. \\
$^{3}$Instituto de Astrof\'{\i}sica de Andalucia, IAA-CSIC, Granada, Spain. \\
$^{4}$Scuola di Ingegneria, Universit\`a degli studi della Basilicata, Potenza, Italy. \\
$^{5}$National Radio Astronomy Observatory, Charlottesville, USA. \\
$^{6}$CSIRO Australia Telescope National Facility, Australia. \\
$^{7}$Max-Planck-Institut f\"ur Radioastronomie, Bonn, Germany. \\
$^{8}$Institute for Astrophysical Research, Boston University, USA. \\
$^{9}$Astronomical Institute, St. Petersburg State University, Russia. \\
$^{10}$Korea Astronomy and Space Science Institute, Republic of Korea. \\
$^{11}$Astro Space Center of Lebedev Physical Institute, Moscow, Russia. \\
$^{12}$Max-Planck-Institut f\"ur Physik, M\"unchen, Germany. \\
$^{13}$Department of Physics and Astronomy, Whittier College, USA. \\
$^{14}$Sternberg Astronomical Institute, Moscow State University, Moscow, Russia.

        E-mail: \email{rocco.lico@unibo.it}
        
               }

\abstract{
In this work we present a polarization analysis at radio frequencies of Markarian 421, one of the closest ($z=0.03$) TeV blazars. 
The observations were obtained, both in total and in polarized intensity, with the Very Long Baseline Array (VLBA) at 15, 24, and 43\,GHz throughout 2011, with one observation per month (for a total of twelve epochs). 
We investigate the magnetic field topology and the polarization structure on parsec scale and their evolution with time. 
We detect polarized emission both in the core and in the jet region, and it varies with frequency, location and time.
In the core region we measure a mean fractional polarization of $\sim1-2\%$, with a peak of about $4\%$ in March at 43\,GHz; the polarization angle is almost stable at 43\,GHz, but it shows significant variability in the range $114^\circ-173^\circ$ at 15\,GHz.
In the jet region the polarization properties show a more stable behavior; the fractional polarization is $\sim16\%$ and the polarization angle is nearly perpendicular to the jet axis. 
The higher EVPA variability observed at 15\,GHz is due both to a variable Faraday rotation effect and to opacity. The residual variability observed in the intrinsic polarization angle, together with the low degree of polarization in the core region, could be explained with the presence of a blend of variable cross-polarized subcomponents within the beam. 

}

\FullConference{12th European VLBI Network Symposium and Users Meeting,\\
		7-10 October 2014\\
		Cagliari, Italy}

\begin{document}

\section{Introduction}

Mrk\,421 is a nearby ($z=0.03$) and bright TeV blazar, for this reason it is a very good candidate for investigating and probing the physical mechanisms occurring in the innermost regions of relativistic jets. 
In general, in these objects the emission is dominated by nonthermal radiation produced by relativistic electrons within the jet, interacting with the magnetic field. Their spectral energy distribution (SED) consists of a low-frequency hump, which represents the synchrotron emission of relativistic electrons within the jet, and a high-frequency hump, that in general is assumed to represent high energy emission produced by inverse Compton (IC) scattering. 
Usually, in these objects, the high-energy IC component is considered as synchrotron-self-Compton emission \citep[SSC, see][]{Abdo2011}, produced from the interaction of the synchrotron photons with the same electrons that produced them. In particular, the synchrotron hump in the SED of Mrk\,421, and of most TeV blazars \citep{Piner2013}, peaks at soft X-rays, and it is classified as a high-synchrotron-peaked (HSP) blazar \citep{Abdo2011}. The analysis of the polarized emission of these objects allow us to obtain information on the magnetic field topology and the emission mechanisms. 

We present Very Long Baseline Array (VLBA) data obtained at 15, 24, and 43\,GHz in total and polarized intensity. We determined important and useful physical parameters such as the absolute orientation of the electric vector position angle (EVPA), the degree of polarization and the Rotation Measure (RM). 
The complete polarization analysis, together with a study of the radio/$\gamma$-ray connection, is presented in \citep{Lico2014}.
We note that at the redshift of the source, 1 mas corresponds to 0.59 pc and that all angles presented in this work are measured from north through east.

\section{Observations details and data reduction}

We observed the target once per month during 2011 with the VLBA at 15, 24, and 43\,GHz, both in total and polarized intensity. The details about the observations at 15 and 24\,GHz are presented in \citep{Lico2012}, and at 43\,GHz in \citep{Blasi2013}.
We expanded our 43\,GHz dataset with 11 additional epochs provided by the VLBA-BU-BLAZAR program\footnote{\url{http://www.bu.edu/blazars/VLBAproject.html}}.
The fringe-fitting, the detection of cross-polarized fringes and the calibration were made by means of the software package Astronomical Image Processing System (AIPS), while the cleaned and final images were produced with the software package DIFMAP.
We adopted the method proposed by \citep{Leppanen1995} to determine the EVPA absolute orientation, based on the instrumental polarization parameters (D-terms). 
For the instrumental polarization calibration we used the calibrator J1310+3220, which is strong (flux density $> 1$ Jy), structureless and has negligible polarization on large scales. 


\begin{figure} 
 \centering
 \includegraphics[bb=0 45 595 667,width=0.32\columnwidth ,clip]{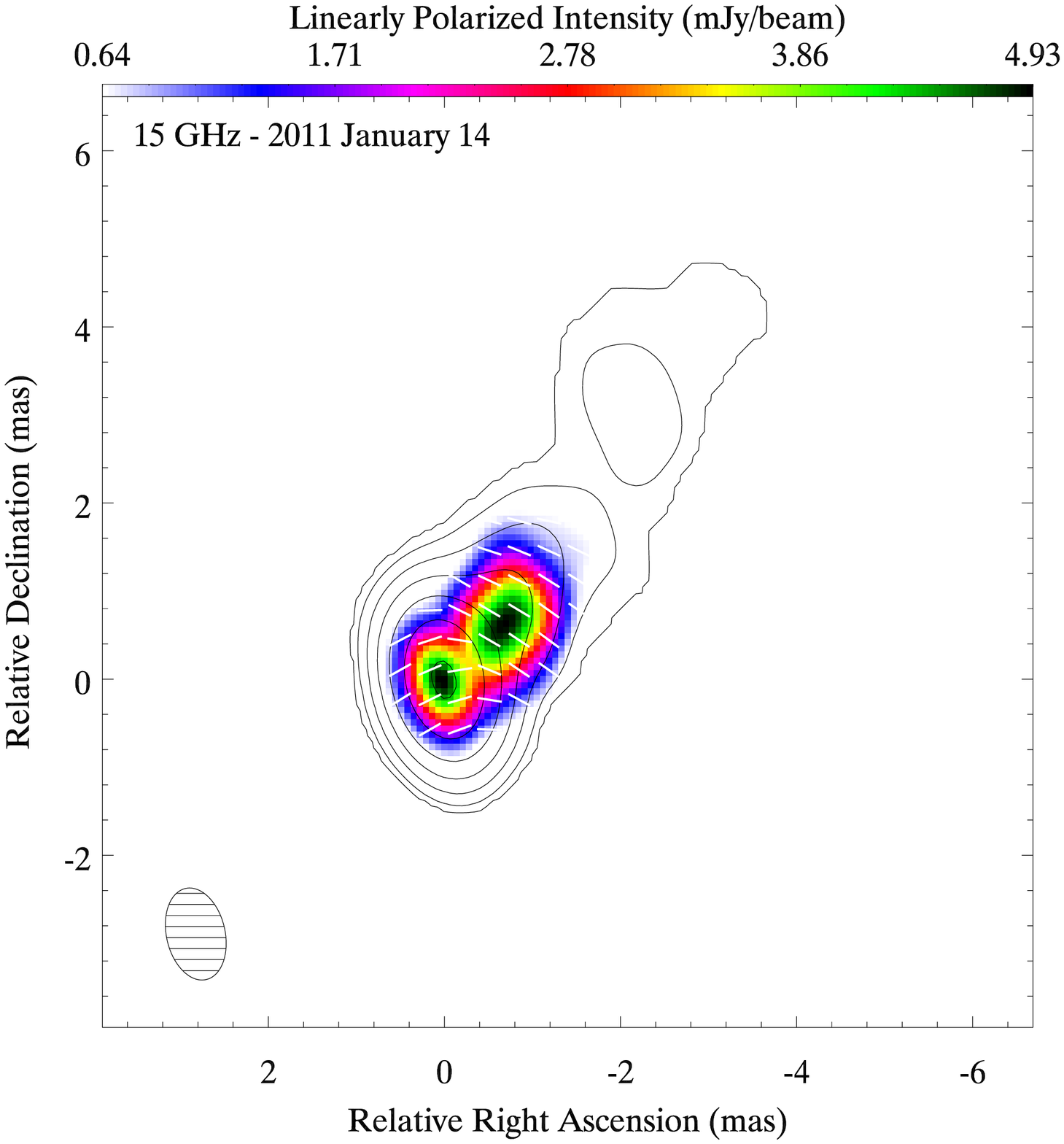}   
 \includegraphics[bb=0 45 595 693,width=0.32\columnwidth ,clip]{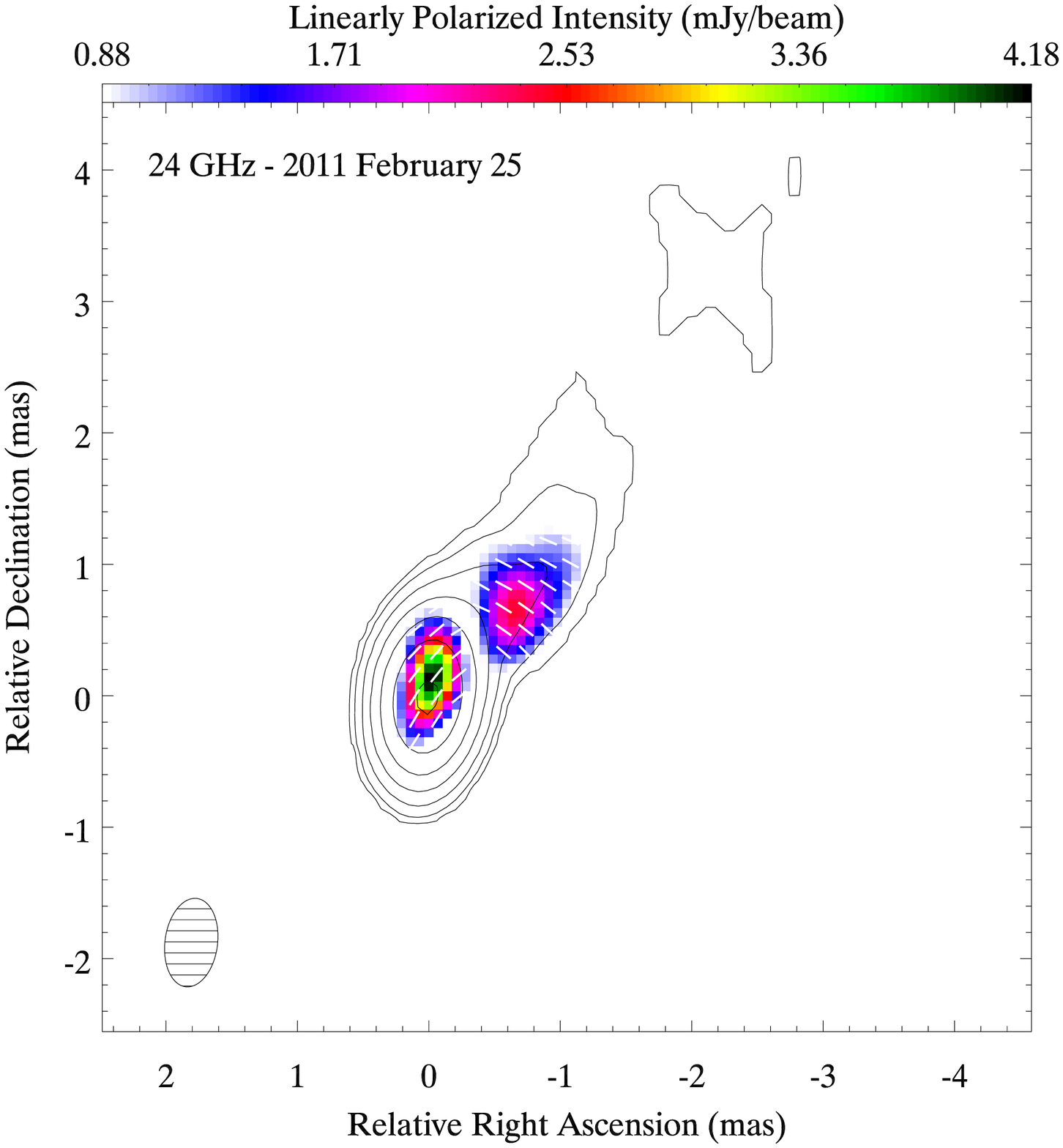} 
 \includegraphics[bb=0 45 595 693,width=0.32\columnwidth ,clip]{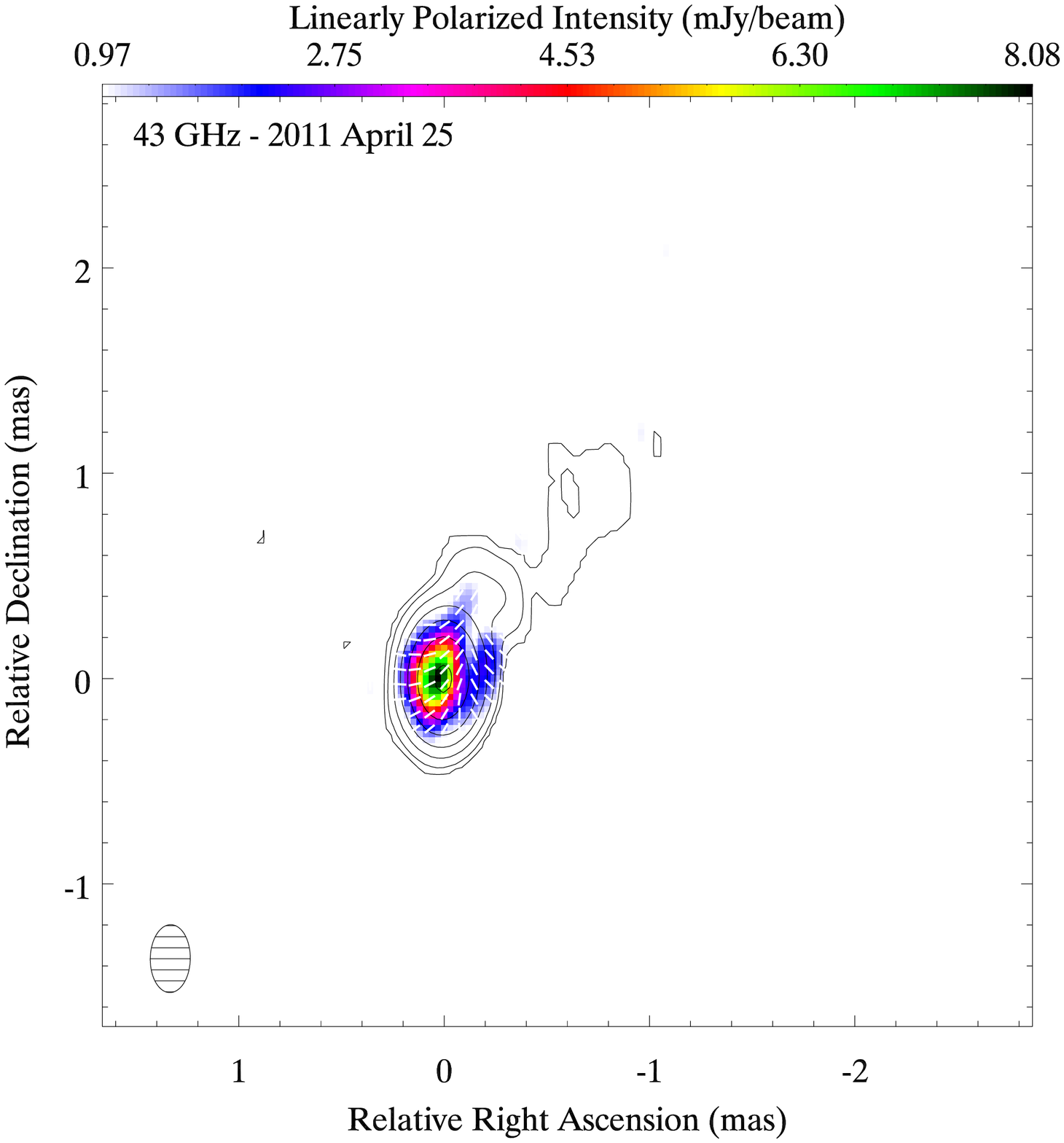} \\
\caption{
{\footnotesize Images of Mrk\,421 at 15\,GHz during the first observing epoch (left image), at 24\,GHz during the second observing epoch (central image), and at 43\,GHz during the third observing epoch (right image). The contour levels are drawn at $(-1, 1, 2, 4...) \times$ the lowest contour (it is at 1.0 mJy/beam for 15 and 24\,GHz and at 0.65 mJy/beam for the 43\,GHz) and increase by factors of 2. In the bottom left corner of each image there is the restoring beam, that is 1.05~mas $\times$ 0.66~mas at 15\,GHz, 0.67~mas $\times$ 0.40~mas at 24\,GHz, and 0.33~mas $\times$ 0.19~mas at 43\,GHz. Colors represent the linearly polarized intensity and the white bars the absolute EVPA orientation.}
}
\label{maps}
\end{figure}

\section{Results}
\subsection{Images and morphology}
A sample of three polarization images of Mrk\,421, produced with DIFMAP and IDL\footnote{\url{http://www.exelisvis.com/ProductsServices/IDL.aspx}}, for each observing frequency, is shown in Fig.~\ref{maps}. All the images are restored with natural weighting to obtain an improved sensitivity for the extended jet emission. In the images the white bars represent the EVPA absolute orientation, the contours the total intensity and the overlaid color maps the linearly polarized intensity. From the total intensity images at all three frequencies, it clearly emerges a collimated and well-defined one-sided jet structure, that extends for $\sim4.5$ mas (2.7 pc) from a compact nuclear region; the position angle (PA) is about $-35^\circ$ (in agreement with \citep{Giroletti2004}).
At 15 and 24\,GHz we detect the linearly polarized emission within a region extending for $\sim1$ mas from the core region, and this allows us to distinguish between core and jet emission. In the outer part of the jet the polarized emission is too faint to be detected. \\
At 43\,GHz, because of sensitivity limitations, only the polarized emission from the core region is detected. In particular, in some epochs we clearly detect a limb brightening structure in the polarized emission in the inner part of the jet. It is clearly revealed in the 4th observing epoch (April 2011), as it is shown in the right image of Fig.~\ref{maps}. 
During this epoch the polarization emission peak in the core region is $\sim8$ mJy/beam, while in the region where the limbs are detected it is $\sim2.3$ mJy/beam.

\begin{figure*}
\centering   
\includegraphics[bb=77 360 539 720, width=0.45\columnwidth ,clip]{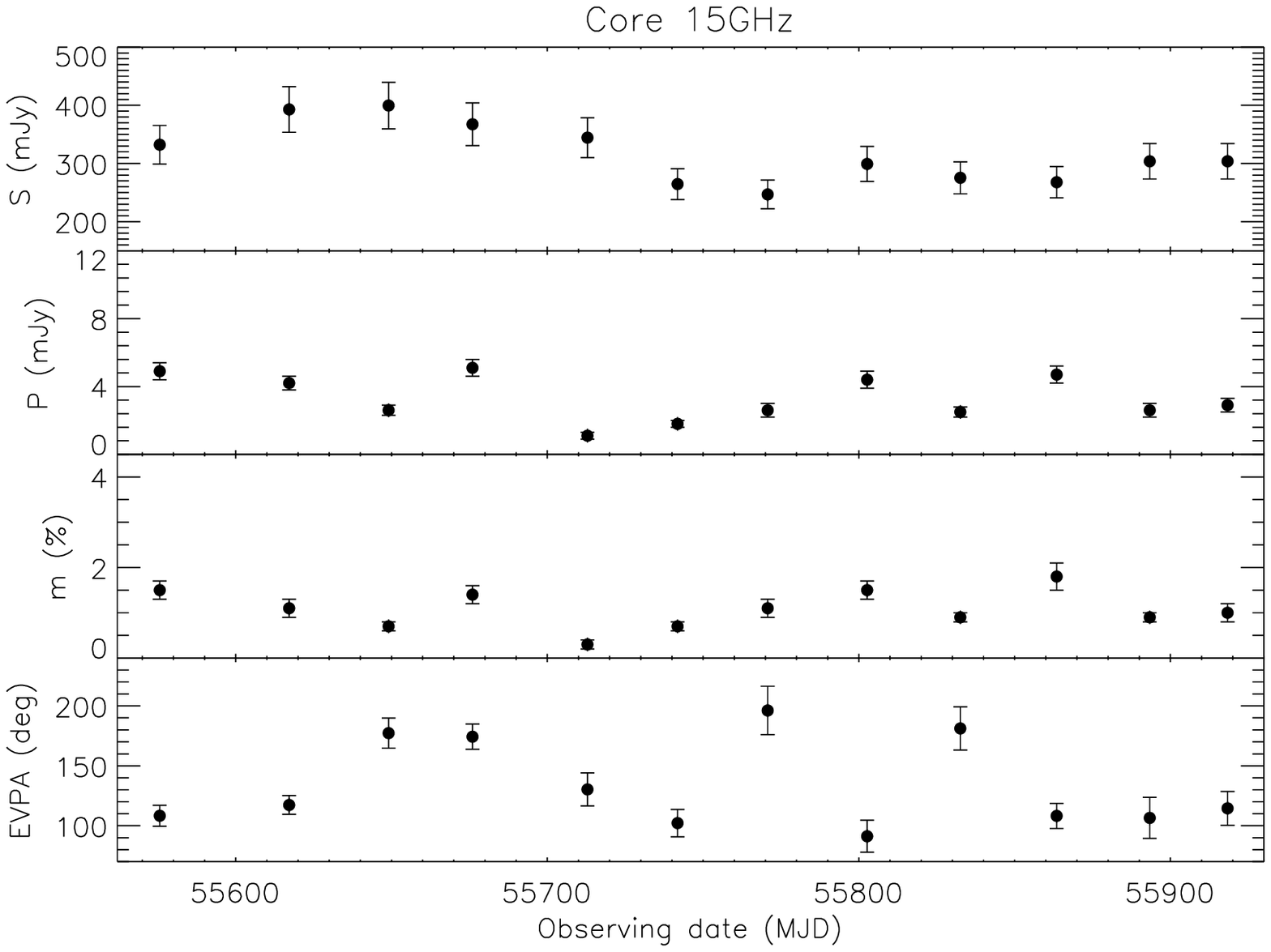}    
\includegraphics[bb=77 360 539 720, width=0.45\columnwidth ,clip]{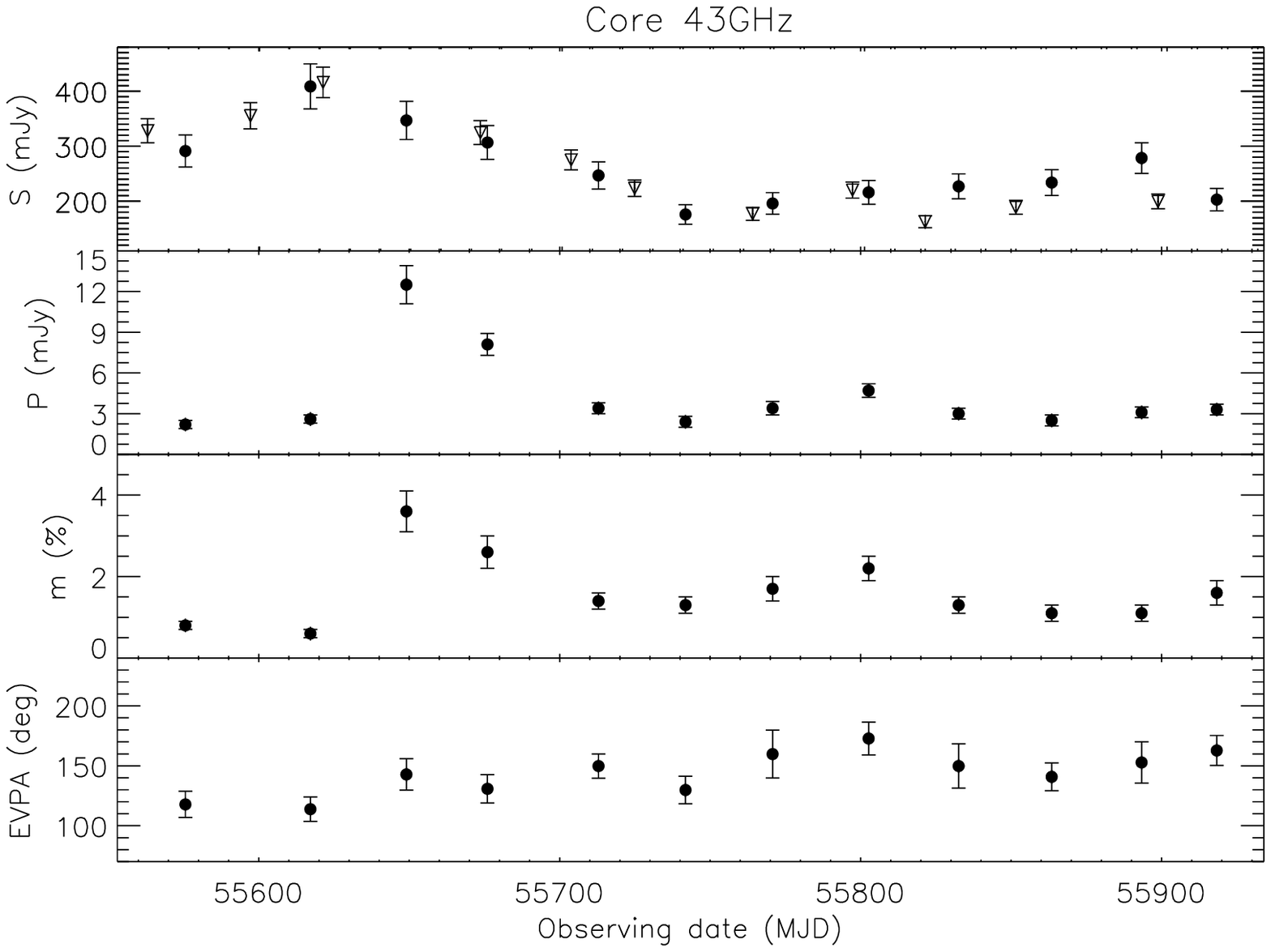} \\
\caption{
{\footnotesize Time evolution in the core region of some physical parameters at 15\,GHz (laft frame) and 43\,GHz (right frame).In each frame from the first to the fourth panel we show the total and polarized flux density light curves, the polarization percentage, and the EVPA values. The additional 43\,GHz VLBA observations (from the VLBA-BU-BLAZAR program) are represented with triangles.}
} 
\label{plots_core}
\end{figure*}

\subsection{Radio light curves and evolution of polarization angle} 

\begin{figure*}
\includegraphics[bb=77 360 539 720, width=0.5\columnwidth ,clip]{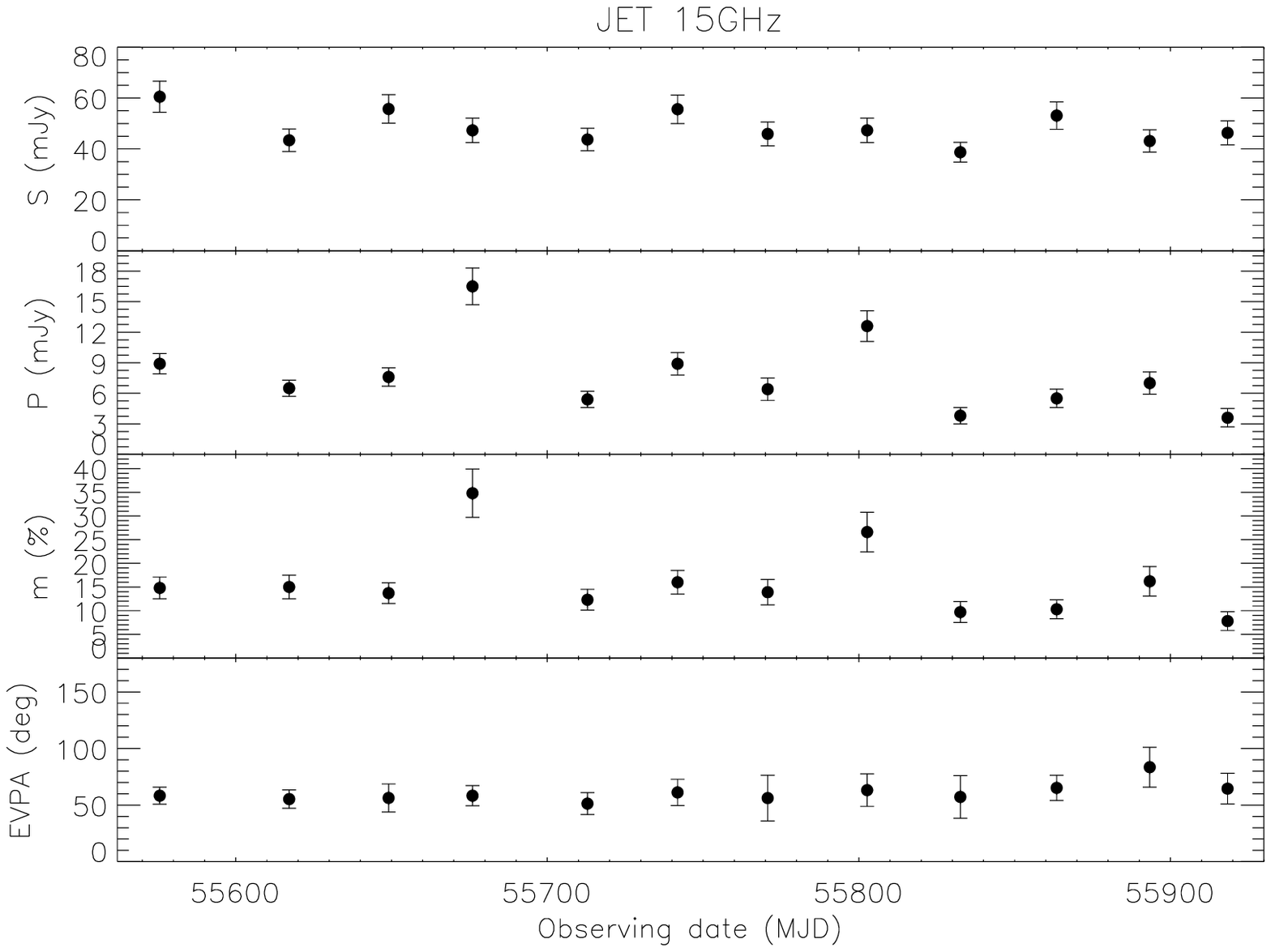} 
\includegraphics[bb=77 360 539 720, width=0.5\columnwidth ,clip]{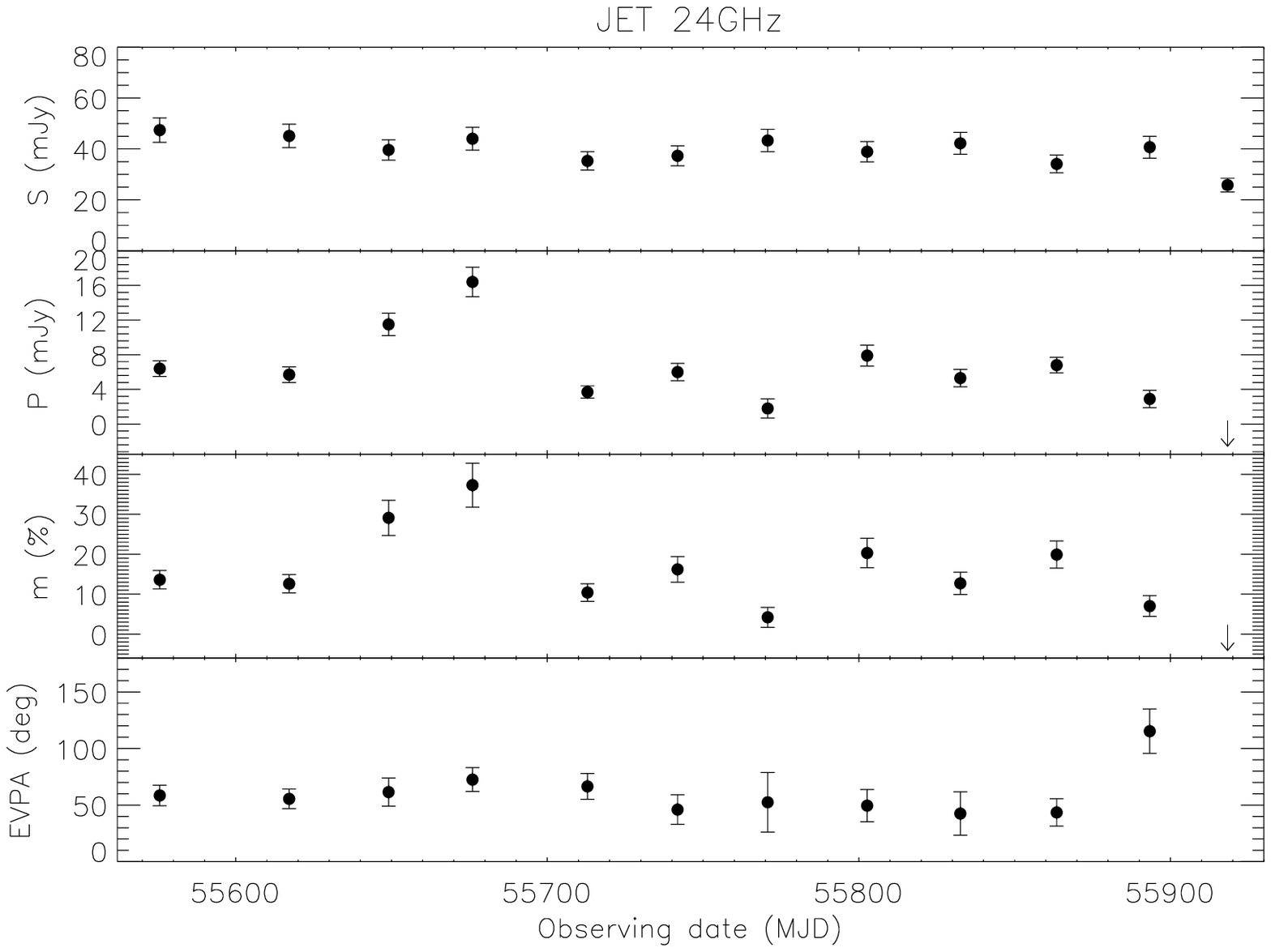} \\
\caption{
{\footnotesize Time evolution of the jet region of some physical parameters at 15\,GHz (left frame) and 24\,GHz (right frame). In each frame from the first to the fourth panel we show the total and polarized flux density light curves, the polarization percentage, and the EVPA values.} 
} 
\label{plots_jet}
\end{figure*}

In the following we will present some fundamental physical parameters both for the core (Fig.~\ref{plots_core}) and the jet (Fig.~\ref{plots_jet}) region. In Fig.~\ref{plots_core} we show two frames, representing 15 (left frame) and 43\,GHz (right frame) data. Each frame has four panels in which we report (from top to bottom) the time evolution of the total and polarized flux density, the polarization percentage, and the EVPAs.
For the jet region we report the evolution of the same parameters in Fig.~\ref{plots_jet} at 15\,GHz (left frame) and 24\,GHz (right frame); at 43\,GHz no polarized emission is detected from the jet. All of the values represented in these images are reported in \citep{Lico2014}.

We start by describing the results for the core region (Fig.~\ref{plots_core}). In the total intensity light curves, for all the three frequencies, we clearly observe a peak around MJD $\sim$55617, followed by a decrease until MJD 55770 and a further slight increase occurring in the last observing period. 
We observe a significantly variable (with a difference between the highest and the lowest value $>3\sigma$) linearly polarized emission both at 15 and 24\,GHz, but the highest variation is observed at 43\,GHz, where we observe a $12.5$ mJy/beam peak during MJD $\sim$55649.
The mean value of the polarization fraction is $\sim1\%$; it is higher at 43\,GHz, with a mean value of $\sim2\%$, and it reaches a peak of $\sim4\%$ during the third observing epoch. The polarization angle varies in the range $110^\circ$-$150^\circ$ for most of the year. In some epochs at 15\,GHz, it shows a variation of $\sim90^\circ$, without any clear connection with the polarized flux density or the EVPA trend both at 24 and 43\,GHz.

In the jet region the situation is very different (Fig.~\ref{plots_jet}). The polarized emission extends for $\sim1$ mas from the core and it is revealed only at 15 and 24\,GHz. In the total intensity light curves we do not observe any significant variation while the polarized flux density is more variable, with a peak during the fourth observing epoch. The fractional polarization is $\sim16\%$. The increasing polarization percentage with the distance from the core is observed in many blazars and it seems to be a common feature  \citep{Lister2001}. The EVPAs show a stable behavior around a value of $60^\circ$, this means that they are roughly perpendicular with respect the jet position angle ($\sim-35^{\circ}$).

\subsection{Faraday rotation analysis}
When a polarized wave propagates through a magnetized plasma, because of the Faraday rotation effect, the observed polarization angle ($\chi_\mathrm{obs}$) is rotated with respect to the intrinsic value ($\chi_\mathrm{int}$), following the relationship $\chi_\mathrm{obs}=\chi_\mathrm{int} + RM\times \lambda^2$, where $\lambda^2$ is the observing wavelength squared and RM is the rotation measure (a quantity related to the electron density $n_e$ and the component of the magnetic field parallel to the line of sight $\textbf{B}_{\parallel}$).
By performing linear fits of EVPAs versus $\lambda^2$ for the core region, where EVPA values at all the three observing frequencies are available, we obtained the RM and $\chi_\mathrm{int}$  values. By addressing the two observed flips in the EVPAs at 15\,GHz (in July and September 2011) to optically thin-thick transitions, we performed the RM fits by rotating them by $90^\circ$. In Fig.~\ref{rm} we show the time evolution of RM (upper panel) and of the intrinsic EVPA values (lower panel). The RM values span from $(-3640 \pm 930)$ to $(+1940\pm750)$ rad\,m$^{-2}$. We note that we can not provide an accurate interpretation of the RM trend because often the uncertainties are very large and many RM values are consistent with 0 within $1\sigma$ or $2\sigma$. The intrinsic polarization angle trend reflects the stable behavior of the 43\,GHz EVPAs, with a mean value of $\sim150^\circ$, i.e. approximately parallel to the jet axis, and it shows some residual variability ($F_{var}$ is $0.10 \pm 0.04$).

\section{Discussion and conclusion}
\begin{figure}
\centering
\includegraphics[width=0.55\columnwidth]{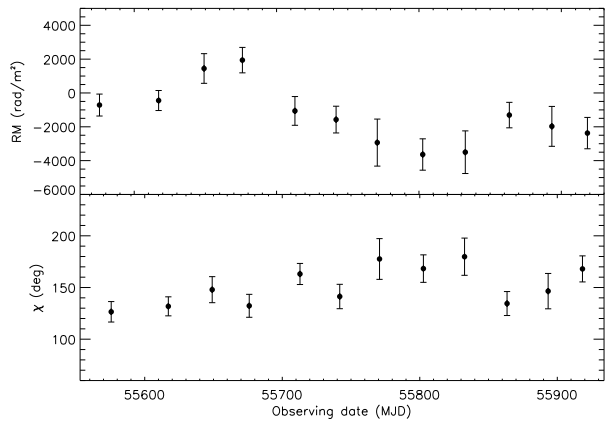} \\
\caption{
{\footnotesize Upper panel: time evolution of the RM values for the core region, obtained by using 15, 24, and 43\,GHz data. Lower panel: time evolution of the polarization angle intrinsic values, obtained from the $\lambda^2$ fits.}
} 
\label{rm}
\end{figure}

In this section we summarize and discuss the polarization properties found for the source, trying to draw a physical scenario to explain them. First, we discuss the EVPA trend observed in different regions of the source. 
In the jet region, we found a stable trend during the entire 2011 for the EVPAs with a mean value of $\sim60^\circ$, i.e. approximately perpendicular to the jet axis. This means that that the magnetic field is parallel to the inner jet position angle; this is unusual in BL Lac objects (see also \citep{Piner2005}). To explain this configuration we can invoke either a helical magnetic field which has a pitch angle that is smaller than $45^\circ$ \citep{Wardle2013} or velocity shear across the jet. 
In the core region the EVPA trend is more complex and it is different from what we observe in the outer jet. At 43\,GHz the EVPAs are approximately parallel to the jet axis with values varying in the range $\sim100^\circ$-$150^\circ$; a similar trend is obtained for the intrinsic polarization angle (lower panel in Fig.~\ref{rm}). This EVPA configuration, if we assume this emission region to be optically thin at 43\,GHz, indicates that the magnetic field is approximately perpendicular to the jet axis. This is in agreement with the presence of a transverse shock. At 15\,GHz, we clearly observe two $90^\circ$ flips of the core EVPAs (from parallel to orthogonal to the jet axis) in July and September. This behavior is the clear signature of an optically thin/thick transition. A possible physical scenario to explain the observed variability at longer wavelengths, might be an association with variable Faraday rotation and opacity effects. The RM variability might be related both to a change in the sign of $\textbf{B}_{\parallel}$ and/or changes in the electron density. Another possibility, that should be taken into account, is the change of the projected path length. 


To explain the residual variability ($F_{var}$ is $0.10 \pm 0.04$) in the intrinsic EVPAs, we invoke the presence of a blend of cross-polarized and variable subcomponents within the beam region; the global polarization properties are determined by their relative contributions which vary as a function of time. This scenario reconciles both the low degree of polarization of the core region (subcomponents with different EVPAs cause significant cancellation) and the residual variations in the intrinsic EVPA values (considering that they are integrated across the VLBI core region).
The presence of the cross-polarized subcomponents, as suggested from the stable EVPA trend at 43\,GHz, can be explained with the following physical scenario: the magnetic field in the core region might be orthogonal to the jet axis, but most part of the emission is optically thick at 15\,GHz, and this gives rise to EVPAs that are orthogonal to the jet axis in the optically thick region and parallel in the optically thin region. 


\end{document}